\documentclass[reprint,secnumarabic,amssymb, showpacs, nobibnotes, aps, prl,10pt,notitlepage]{revtex4-1}

\usepackage{graphicx}% Include figure files
\usepackage{graphics}
\usepackage{dcolumn}% Align table columns on decimal point
\usepackage{bm}% bold math
\usepackage{multirow}
\usepackage{xspace}
\usepackage{amsmath}
\usepackage{calrsfs}
\usepackage{braket}
\usepackage{hyperref}
\hypersetup{colorlinks=true, linkcolor=blue, citecolor=blue, urlcolor=blue, unicode=false}
\newcommand{\Si}{Si(111)-($7\times 7$)\xspace}

\newcommand{\K}{$\bar{\text{K}}$\xspace}
\newcommand{\M}{$\bar{\text{M}}$\xspace}
\newcommand{\G}{$\bar{\Gamma}$\xspace}
\newcommand{\KMK}{$\bar{\text{K}}-\bar{\text{M}}-\bar{\text{K}}$\xspace}
\newcommand{\GK}{$\bar{\Gamma}-\bar{\text{K}}$\xspace}
\newcommand{\GM}{$\bar{\Gamma}-\bar{\text{M}}$\xspace}
\newcommand{\SP}{S1$^+$\xspace}
\newcommand{\SN}{S1$^-$\xspace}

\begin{document}

\title{A reinvestigation of the giant Rashba-split states on Bi-covered Si(111)}

\author{M.~H.~Berntsen}
\email{mhbe@kth.se}
\affiliation{KTH Royal Institute of Technology, SCI Materials Physics, S-164 40 Kista, Sweden}
\author{O.~G\"{o}tberg}
\affiliation{KTH Royal Institute of Technology, SCI Materials Physics, S-164 40 Kista, Sweden}
\author{O.~Tjernberg}
\email{oscar@kth.se}
\affiliation{KTH Royal Institute of Technology, SCI Materials Physics, S-164 40 Kista, Sweden}

\date{\today}

\begin{abstract}
We study the electronic and spin structures of the giant Rashba-split surface states of the Bi/Si(111)-$(\sqrt{3}\times\sqrt{3}$)R$30^{\circ}$ trimer phase by means of spin- and angle-resolved photoelectron spectroscopy (spin-ARPES). Supported by tight-binding calculations of the surface state dispersion and spin orientation, our findings show that the spin experiences a vortex-like structure around the $\bar{\Gamma}$-point of the surface Brillouin zone --- in accordance with the standard Rashba model. Moreover, we find no evidence of a spin vortex around the $\bar{\mathrm{K}}$-point in the hexagonal Brillouin zone, and thus no peculiar Rashba split around this point, something that has been suggested by previous works. Rather the opposite, our results show that the spin structure around \K can be fully understood by taking into account the symmetry of the Brillouin zone and the intersection of spin vortices centered around the $\bar{\Gamma}$-points in neighboring Brillouin zones. As a result, the spin structure is consistently explained within the standard framework of the Rashba model although the spin-polarized surface states experience a more complex dispersion compared to free-electron like parabolic states.

\end{abstract}

% 73.20.-r 		Electron states at surfaces and interfaces
% 73.20.At		Surface states, band structure, electron density of states
% 71.70.Ej		Spin-orbit coupling, Zeeman and Stark splitting, Jahn-Teller effect
% 79.60.-i		Photoemission and photoelectron spectra 
% 74.20.Pq		Electronic structure calculations 

\maketitle

\section{Introduction}
When the inversion symmetry of a crystal is broken, spin-orbit interaction can lift the spin degeneracy leading to the appearance of spin-polarized electronic states in momentum space. For two-dimensional electron gases (2DEG) at surfaces or interfaces, the phenomenon is commonly referred to as the Rashba effect~\cite{Bychkov1984}, and the existence of spin-split surface states with a close to ideal free-electron-like dispersion has been confirmed on noble metallic surfaces~\cite{LaShell1996,Nicolay2001,Hoesch2004,Reinert2001}. Other two-dimensional systems with more complex band structures can also display spin split states~\cite{Hochstrasser2002,Koroteev2004,Sugawara2006,Bentmann2009,Ast2007,Ast2008}, in some cases with a sizable energy separation between the spin branches. 

The technological interest in systems hosting spin-split electronic states is rooted in the idea of exploiting the spin properties of electrons for information transfer in spintronic devices~\cite{Datta1990,Wolf2001}. As an alternative to spin injection through ferromagnetic layers, there has been an increased interest towards direct generation of spin-polarized currents using semiconductor devices with spin-polarized states. For this purpose, spin states with a large energy separation is required to efficiently, and unequivocally, separate the two spins. 

In this context, one specific system that has received attention in the past is the $\mathrm{\beta}$-Bi surface, which exhibits a giant spin-split in the order of several hundred milli-electron volts (meV). The $\mathrm{\beta}$-Bi surface is formed by adsorbing a single monolayer (ML) of Bi onto a Si(111) surface and can therefore be viewed as a strictly two-dimensional system. The large spin split in this system has been subject to previous investigations by angle-resolved photoelectron spectroscopy (ARPES) --- including spin-resolved photoelectron spectroscopy (SR-PES)~\cite{Kim2001,Sakamoto2009,Gierz2009,Frantzeskakis2010}. Spin-resolved band structure calculations have also confirmed the spin polarization of the surface states~\cite{Sakamoto2009,Gierz2009,Frantzeskakis2010}. These studies discovered a non-trivial band and spin structure in the vicinity of the \K and \M points of the hexagonal surface Brillouin zone (SBZ), indicating that time-reversal symmetry is not a necessary condition or a guarantee for experiencing a simple vortical spin structure around a point in the SBZ --- as predicted by the standard Rashba model. In spite of the efforts of previous works, there are still open questions regarding the detailed evolution of both the band structure and the spin texture around the \K and \M points in this particular system. For instance, some inconsistencies between experimental data and theoretical band-structure calculations exist in literature, which have consequences for the interpretation of the presence, or the absence of, a vortical spin structure around the \K and \M points. 

Here, we present new spin- and angle-resolved PES data on the $\mathrm{\beta}$-Bi system together with renewed band structure calculations from an optimized tight-binding (TB) model with the aim of arriving at a more accurate description of the band and spin structures in the $\mathrm{\beta}$-Bi phase. Supported by our calculations, we are able to follow the evolution of experimental constant energy contours around the \K and \M points for different binding energies, revealing changes in the topology of the contours caused by the presence of several saddle points in the band structure. Furthermore, we show that in spite of exhibiting a complex, vortical, band structure around the \K-point the overall spin behavior around this point in the SBZ follows naturally from a simpler, circulating spin structure around \G. As a consequence, the peculiar Rashba split around \K suggested by earlier works is in fact shown to be rather nonpeculiar. In the same way, we confirm the nonvortical behavior of the spin around the \M-point. Additionally, our optimized tight-binding calculations correctly positions the extremal points of the band with highest energy along the \GM direction, in agreement with our data, thereby resolving a discrepancy between past calculations and experiments.

\section{Experimental details}
\begin{figure}
\includegraphics[width=8.6 cm]{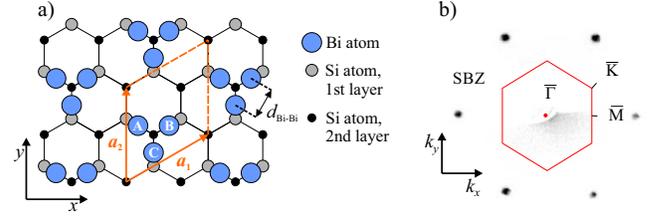}
\caption{Schematic drawing of the trimer structure of the $\mathrm{\beta}$-Bi phase on Si(111). The lattice vectors, $\bm{a}_1$ and $\bm{a}_2$, of the Bi-structure and the Bi-Bi distance within the trimer, $d_{\text{Bi-Bi}}$, are indicated in the figure. The atoms in the trimer are labeled A, B and C. b) An experimental LEED image from the ($\sqrt{3}\times\sqrt{3}$)R$30^\circ$ structure of $\mathrm{\beta}$-Bi together with the corresponding SBZ (red hexagon).}
\label{structurefig}
\end{figure}

The monolayer Bi $\beta$-phase on Si(111) consists of a trimer structure of Bi atoms centered above a Si atom from the second surface layer in the $T_4$ position~\cite{Wan1991,Wan1992,Shioda1993,Cheng1997,Miwa2003,Kuzumaki2010}, as schematically depicted in Fig.~\ref{structurefig}a). The Bi trimers form a $(\sqrt{3}\times\sqrt{3}$)R$30^{\circ}$ superstructure and result in a hexagonal surface Brillouin zone (SBZ) as drawn in Fig.~\ref{structurefig}b), overlaid on an experimental low-energy electron diffraction (LEED) pattern from the present study. The $\mathrm{\beta}$-Bi surface used in this work was produced in-situ by deposition of 1~ML Bi onto a freshly prepared \Si surface using e-beam evaporation. The reconstructed Si surface was prepared by annealing of the Si(111) substrate (arsenic doped, resistivity 4~m$\Omega$cm) to 1100~$^{\circ}$C by direct current heating. A quartz crystal microbalance was used to calibrate the deposition rate prior to the Bi deposition and during the growth of the monolayer the Si-substrate temperature was kept at 470~$^{\circ}$C. The final structure of the $(\sqrt{3}\times\sqrt{3}$)R$30^{\circ}$ surface was verified by LEED. During growth, the base pressure in the deposition chamber stayed below $5\cdot10^{-10}$~mbar.

Spin- and angle-resolved photoelectron spectroscopy (spin-ARPES) measurements were carried out with linearly polarized light ($h\nu = 25$~eV) using a spin- and angle-resolving photoemission setup~\cite{Berntsen2010} at the I3 beamline on the MAXIII storage ring at MAXlab, Lund, Sweden. All photoemission spectra were acquired at room temperature and at a base pressure below $2\cdot10^{-10}$~mbar. The energy and angular resolutions of the photoelectron analyzer were approximately 10~meV and 0.1~degrees, respectively, for the angle resolved measurements. For the spin resolved spectra, the corresponding resolutions were 100~meV and 3~degrees, respectively. 

\section{Theoretical model}
For modeling of the measured band structure, we use a similar tight-binding (TB) model as introduced by Frantzeskakis \textit{et al}.~\cite{Frantzeskakis2010}. The model assumes a $sp_z$ character of the dominating orbital, thus allowing for a parametrization of the overlap integrals as a function of the distance between neighboring atoms without any angular dependence. The tight-binding Hamiltonian including spin-orbit (SO) coupling is given in a site representation by
\begin{multline}
 H_{\text{TB}}=H_{\text{0}}+H_{\text{SO}}=-\sum_{\bm{n},\bm{m},\alpha,\beta,i,j} t_{\alpha\beta}(\bm{m})\text{a}^{\dagger}_{\alpha(\bm{n}+\bm{m})i}\text{a}_{\beta\bm{n}j} \\
 +\mathrm{i}\sum_{\bm{n},\bm{m},\alpha,\beta,i,j}\lambda_{\alpha\beta}(\bm{m})\text{a}^{\dagger}_{\alpha(\bm{n}+\bm{m})i} \left(\bm{\sigma}\times \bm{\hat{d}}_{\alpha\beta}(\bm{m})\right)_{z} \text{a}_{\beta\bm{n}j}.
 \label{SiteTB}
\end{multline}
Here, $\bm{n}$ is a vector running over all unit cells and $\bm{m}$ a vector for an atom's nearest neighbors. $\alpha$ and $\beta$ index the three trimer atoms (A, B and C) in the base. $t_{\alpha\beta}(\bm{m})$ is the overlap integral between atom $\alpha$ and the nearest neighbor $\beta$ at $\bm{m}$. $\lambda_{\alpha\beta}(\bm{m})$ is the corresponding SO coefficient and $\bm{\hat{d}}_{\alpha\beta}(\bm{m})$ the unit vector along the direction to the nearest neighbor $\bm{m}$. $\bm{\sigma}$ are the Pauli spin matrices and the sub-index $z$ indicates the $z$-component of the cross product. $\text{a}^{\dagger}_{\alpha\bm{n}i}$($\text{a}_{\alpha\bm{n}i}$) is the creation (annihilation) operator of an electron with spin state $i$ in the atomic site $\alpha$ at $\bm{n}$. We use a power law dependence for the overlap integrals and the SO coefficients on the form $a\cdot d^{-b}$, where $d$ is the distance between site $\alpha$ and $\beta$ and $a$ and $b$ are independent sets of parameters for the two cases, respectively. Using a basis consisting of spin-up and spin-down states from each trimer site, $\ket{\alpha~i}\in\left\{ \ket{A\uparrow}, \ket{A\downarrow}, \ket{B\uparrow}, \ket{B\downarrow}, \ket{C\uparrow}, \ket{C\downarrow} \right\}$, the momentum representation of the Hamiltonian can be expressed as a 3~$\times$~3 matrix having matrix elements ($\mathcal{H}_{\alpha\beta}$) given by the $2\times 2$ sub-matrices
\begin{multline}
\mathcal{H}_{\alpha\beta}=-I_2\sum_{\bm{m}} t_{\alpha\beta}(\bm{m})   \mathrm{e}^{-\mathrm{i}\bm{kT_m}} \\
 +\mathrm{i}\sum_{\bm{m}} \lambda_{\alpha\beta}(\bm{m}) \mathrm{e}^{-\mathrm{i}\bm{kT_m}} \left(\bm{\sigma}\times \bm{\hat{d}}_{\alpha\beta}(\bm{m})\right)_{z},
 \label{MomentumTB}
\end{multline}
thus accounting for the two spin directions. In Eq.~(\ref{MomentumTB}), $\alpha$ and $\beta$ again runs over the three trimer atoms (A, B and C), $I_2$ is the $2\times2$ identity matrix, $\bm{T_m}$ is the real space vector from a given atom to a neighbor at $\bm{m}$ and $\bm{k}$ a two-dimensional momentum space vector. The $\bm{k}$-dependence of the eigenvalues ($\varepsilon_{l}$) of the Hamiltonian matrix yields the energy dispersion $E(\bm{k})$ of the three spin-polarized bands in our model (in total six spin branches). The corresponding eigenvectors ($\ket{l,\bm{k}}$) can be used to calculate the expectation value of the spin along the $x$, $y$ and $z$ directions for a spin branch $l$ and a point $\bm{k}$ in momentum space as $\braket{S_{\text{n}}(\bm{k})}_{l}=\hbar/2\braket{l,\bm{k}|\sigma_{\text{n}}|l,\bm{k}}$, where $\text{n}=x,y,z$ and $\sigma_{\text{n}}$ denote the corresponding Pauli spin matrix. The TB model is evaluated including fourth order nearest neighbors.

In the context of spin-orbit coupling and spin-split bands it is also of interest to calculate the Berry curvature in the same model. Here, the out of plane component of the Berry curvature is evaluated as a sum over the eigenstates and can be expressed as~\cite{Xiao2010,Gradhand2012}
\begin{equation}
\Omega^l_z(\bm{k}) = \Omega^l_1(\bm{k}) + \Omega^l_2(\bm{k}), 
\label{BCmain}
\end{equation}
where
\small
\begin{equation}
%\scalebox{1}[1]
\Omega^l_1(\bm{k}) = -\mathrm{Im} \sum_{l\neq l'} \frac{\bra{l,\bm{k}}\nabla_{\bm{k}}\mathcal{H}(\bm{k})\ket{l',\bm{k}}\times \bra{l',\bm{k}}\nabla_{\bm{k}}\mathcal{H}(\bm{k})\ket{l,\bm{k}} }{(\varepsilon_l(\bm{k}) - \varepsilon_{l'}(\bm{k}))^2} 
\label{BC1}
\end{equation}
\normalsize
and
\small
\begin{equation}
\Omega^l_2(\bm{k}) =\sum_{l\neq l'} 2 \mathrm{Re} \left[ \frac{\bra{l,\bm{k}}\nabla_{\bm{k}}\mathcal{H}(\bm{k})\ket{l',\bm{k}}\times \bra{l',\bm{k}}\bar{\bar{\bm{\mathrm{r}}}}\ket{l,\bm{k}} }{\varepsilon_l(\bm{k}) - \varepsilon_{l'}(\bm{k})} \right].
\label{BC2}
\end{equation}
\normalsize
Here $\mathcal{H}$ is the full Hamiltonian matrix constructed using Eq.~(\ref{MomentumTB}) and $\bar{\bar{\bm{\mathrm{r}}}}$ is a vector-valued matrix with matrix elements given by $\bm{\mathrm{r}}_{\alpha\beta}=\braket{\beta~i|\bm{\mathrm{r}}|\alpha~i}$, where $\ket{\alpha~i}$ is our site-localized basis functions, see ref.~\onlinecite{Gradhand2012} for further details. The first term in Eq.~(\ref{BCmain}) is identical to the expression found in ref.~\onlinecite{Xiao2010} for determining the Berry curvature using Bloch-state wavefunctions. However, since our $\ket{l,\bm{k}}$ are the eigenvectors of the Hamiltonian matrix, and not the cell-periodic part of the Bloch state, we need to add a dipole correction term, which enters as the second term in Eq.~(\ref{BCmain}).

When optimizing the parameters in the TB model, we evaluate the qualitative agreement of the calculated band structure with our experimental data. Furthermore, we focus at the two, in energy, highest lying spin bands, which correspond to the bands in our experimental data with lowest binding energy. A consequence of our optimization is that we ultimately deviate from a precise physical description of some of the model parameters. For instance, the optimized distance between Bi atoms within the trimer is 30~\% to 40~\% smaller in our model compared to values found in literature~\cite{Wan1991,Wan1992,Cheng1997,Miwa2003,Kuzumaki2010}. Yet, we choose to accept this deviation since the qualitative agreement with the experimental data is improved compared to when the Bi-Bi distance is set to a fixed, physically accurate, value. By changing how the overlap integral and SO coefficients are parametrized we could essentially arrive at a similar optimized band structure even with a realistic value for Bi-Bi distance. However, the possibility for direct comparison of our calculations with those presented by Frantzeskakis \textit{et al}. motivates us to keep the current parametrization.  An overview of the optimized parameters of our model is listed in Table~\ref{table1}. 
\begin{table}
\begin{tabular*}{\columnwidth}{@{\extracolsep{\fill}}  lcc}
\multicolumn{3}{c}{}   \\ \hline \hline
TB parameter & Value present study & Value from [17]\\ \hline
$l_a$ (\AA) & 6.7 & 6.7 \\
$d_{\text{Bi-Bi}}$ (\AA) & 1.8 & 2.6 \\
$a_t$\footnotemark[1] & -2.94 & -2.94 \\
$b_t$\footnotemark[2] & 1.13 & 1.13 \\
$a_\lambda$\footnotemark[3] & 0.40 & 0.15 \\
$b_\lambda$\footnotemark[4] & 1.8 & 0.80 \\ \hline \hline
\end{tabular*}
\footnotetext[1]{prefactor in $t_{\alpha\beta}$}
\footnotetext[2]{exponent in $t_{\alpha\beta}$}
\footnotetext[3]{prefactor in $\lambda_{\alpha\beta}$}
\footnotetext[4]{exponent in $\lambda_{\alpha\beta}$}
\caption{Parameter values for the optimized tight-binding model.}
\label{table1}
\end{table}

\section{Results and Discussions}
\begin{figure*}
\includegraphics[width=\textwidth]{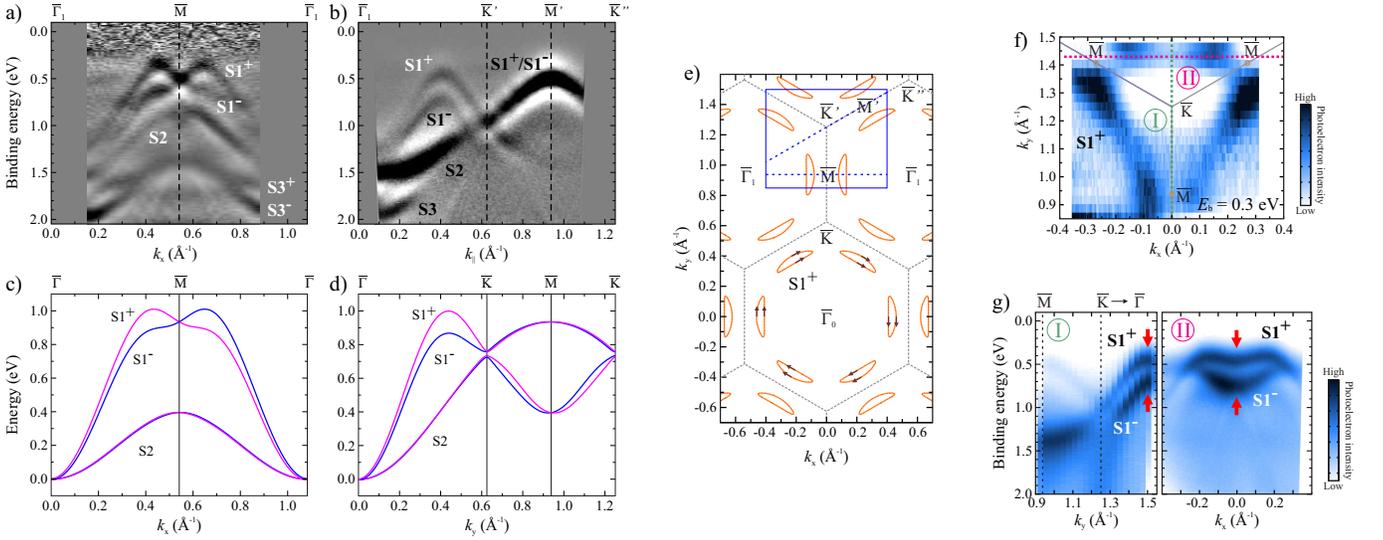}
\caption{a)--b) ARPES second derivative plots along symmetry directions in the SBZ. c)--d) Calculated spin-polarized band structure along the same symmetry lines as in a) and b). e) Calculated constant energy contours for an energy close to the maximum of the \SP branch. The blue rectangle (inset) shows the region in $k$-space where the experimental data has been acquired. The dashed lines show where the data in a) and b) have been extracted. f) Experimental constant energy surface around the \K-point at a binding energy close to the maximum of the \SP branch. g) Band dispersion along the dashed lines I and II in f). Red arrows mark the positions of the saddle points of the \SP and \SN branches.}
\label{datafig}
\end{figure*} 
The experimentally determined band structure of $\mathrm{\beta}$-Bi is presented in Fig.~\ref{datafig}a) and \ref{datafig}b) as second derivative plots of the photoelectron intensity measured along selected symmetry directions in the SBZ. The photoemission data are acquired in the second and third Brillouin zones, as indicated by the blue dashed lines in Fig.~\ref{datafig}e). 

Along the \GM and \GK directions we observe three bands, which appear to split into sub-branches at different positions along these momentum directions. The data are in excellent agreement with the band structure measured in ref.~\onlinecite{Sakamoto2009} from which we conclude that the observed bands in our study are surface states associated with the Bi/Si(111)-($\sqrt{3}\times\sqrt{3}$)R$30^{\circ}$ trimer phase. Furthermore, we adopt a similar labeling of the surface states S1, S2 and S3, as seen from Figs.~\ref{datafig}a) and \ref{datafig}b), noting that the sub-branches of a band are labeled with and additional ``$+$'' for the upper branch (lower binding energy in data) and a ``$-$'' for the lower branch (higher binding energy in data). Such a labeling is unique only within a given Brillouin zone, when taking into account the spin direction of the individual branches. Nonetheless, whenever a band crosses a zone boundary to a neighboring zone, we continue to label the upper branch with a ``$+$'' and the lower with a ``$-$''. 

Focusing on the S1 band in Fig.~\ref{datafig}a), we see that the \SP and \SN branches are degenerate at the \M-point but split in energy elsewhere along the \GM direction. When approaching \G, the energy split decreases. Along the \GK direction, seen in Fig.~\ref{datafig}b), S1 is also split into two branches that become degenerate at \K and which seem to merge when approaching \G. In contrast, no clear energy split in S1 is discernible in our data along the \KMK direction.
 
Investigating in detail the binding energy of the maxima of the \SP branch along \GM and \GK reveals that the global maximum of \SP is along the \GM direction. A constant energy (CE) map of an area of the SBZ covering a \K-point and the surrounding \M points, taken at a binding energy of $E_{\text{b}} = 0.30$~eV, confirms the location of the extrema of the \SP branch, as displayed in Fig.~\ref{datafig}f). The CE map is acquired over the region in $k$-space enclosed by the blue rectangle in Fig.~\ref{datafig}e). Angle resolved cuts of the band structure in $k_x$ and $k_y$ directions, indicated by the dashed lines I and II in Fig.~\ref{datafig}f), furthermore reveals that the \SP branch experiences a saddle point along \GK, at the position where the branch has its maximum along that direction. This saddle point is responsible for the disconnected intensity pockets that appear outside \M, as seen in Fig.~\ref{datafig}f), when looking at a CE map intersecting the \SP branch close to its maximum. 

Going back to the \SN branch in Fig.~\ref{datafig}b), we observe that it too has a maximum along \GK, similar to \SP. However, along the \GM direction, displayed in Fig.~\ref{datafig}a), the maximum occurs only at the Brillouin zone boundary --- at the \M-point. Nevertheless, the maximum of \SN lies higher in energy (at lower binding energy) along \GM compared to the \GK direction. This tells us that also the \SN branch has a saddle point in its ($k_x$,$k_y$,$E_{\text{b}}$)-dispersion --- appearing along \GK. That this is so, we see from the two panels in Fig.~\ref{datafig}g) in a similar way as for \SP. In the same plots, the saddle points of the two S1 branches are marked with arrows.
 
Given the dispersion of the \SP/\SN branches along \KMK and the knowledge of where the two branches have their global maxima reveals an important difference between \SP and \SN around \M. Since the former has its global maximum away from \M, but also a local maximum at \M along the \KMK direction, \SP has a second saddle point located at \M. In contrast, \SN has its global maximum at \M and thus does not experience a saddle point there. Consequently, when moving from the Fermi level ($E_{\text{F}}$) towards higher binding energies, disconnected CE contours for \SP will first develop outside \M whereas the contours from \SN will develop enclosing the \M-point. Sorted in energy by descending order (by increasing binding energy), the saddle point of \SP located outside \K is the first to be reached, then the one of \SP at \M, and lastly the one of \SN located outside \K. We lable these saddle points SP$_1$, SP$_2$ and SP$_3$, respectively.

Looking at the S2 and S3 bands in Figs.~\ref{datafig}a) and \ref{datafig}b), their dispersion is less visible compared to the S1 band. For S2, there is no clearly visible energy split apart from in the near vicinity of the \M-point. The S3 band has a crossing of the S3$^+$ and S3$^-$ branches at \M and an increasing energy split between the sub-branches when approaching \G along \GM.
 
Turning to our TB model, the results of the calculations are displayed in Figs.~\ref{datafig}c) and \ref{datafig}d), and show the spin-resolved dispersion of the S1 and S2 bands along high symmetry directions corresponding to the experimental plots in panels a) and b) of the same figure. When determining the parameters in the TB model, we paid particular attention to the location of the global maxima of the \SP and \SN branches, optimizing the TB parameters such that the calculated band structure was in qualitative agreement with our experimental observations of the S1 band. Consequently, our model less accurately describes the dispersion of the S2 and S3 bands visible in the data. In fact, even though not shown in Figs.~\ref{datafig}c) and \ref{datafig}d), the S3 band also exists in the calculations but appears at around $-9$~eV (at \G) on the calculated energy scale and is for the sake of clarity omitted from the plots. Due to the discrepancy between data and model for the S2 and S3 bands, in the remainder of our work, we will focus only on the S1 band.
 
Interestingly, when selecting realistic values of the interatomic distance between Bi atoms in the trimer, the TB model fails to describe the measured band structure in three important aspects. Firstly, the maximum of the \SN branch along \GM does no longer occur at the \M-point, but rather slightly before \M~--- similar to the \SP branch. Secondly, crossings of the \SP and \SN branches occur within the SBZ, both along \GM and \GK. Thirdly, the global maxima of \SP and \SN move so that they now appear along the \GK direction. This means that CE contours of the \SP branch firs develop outside \K, but more importantly, that \M becomes a saddle point also for the \SN branch. This behavior matches the calculations presented in ref.~\onlinecite{Frantzeskakis2010}, but clearly is not in agreement with our experimental data, as seen from Figs.~\ref{datafig}a)--b) and \ref{datafig}f)--g). Relaxing the constraint of maintaining realistic values for the Bi-Bi distance --- c.f. Table~\ref{table1} --- allows us to arrive at a calculated band structure that is in better agreement with the dispersion of the S1 band. The saddle points of the \SP and \SN branches outside \K are reproduced, as is the saddle point of \SP at \M. For the \SN branch, the \M-point correctly becomes the global maximum. Figure~\ref{datafig}e) shows that the optimized TB model yields CE contours for energies close to the maximum of \SP in the correct position outside the \M-point, c.f. data in Fig.~\ref{datafig}f). Calculating the spin components for points on these contours reveals an in-plane spin that rotates in a clockwise fashion around \G, as indicated by the arrows in Fig.~\ref{datafig}e).  Furthermore, the model yields an opposite spin polarization of the \SP and \SN branches along \GM and \GK, as expected for a spin split band. 
\begin{figure*}
\includegraphics[width=\textwidth]{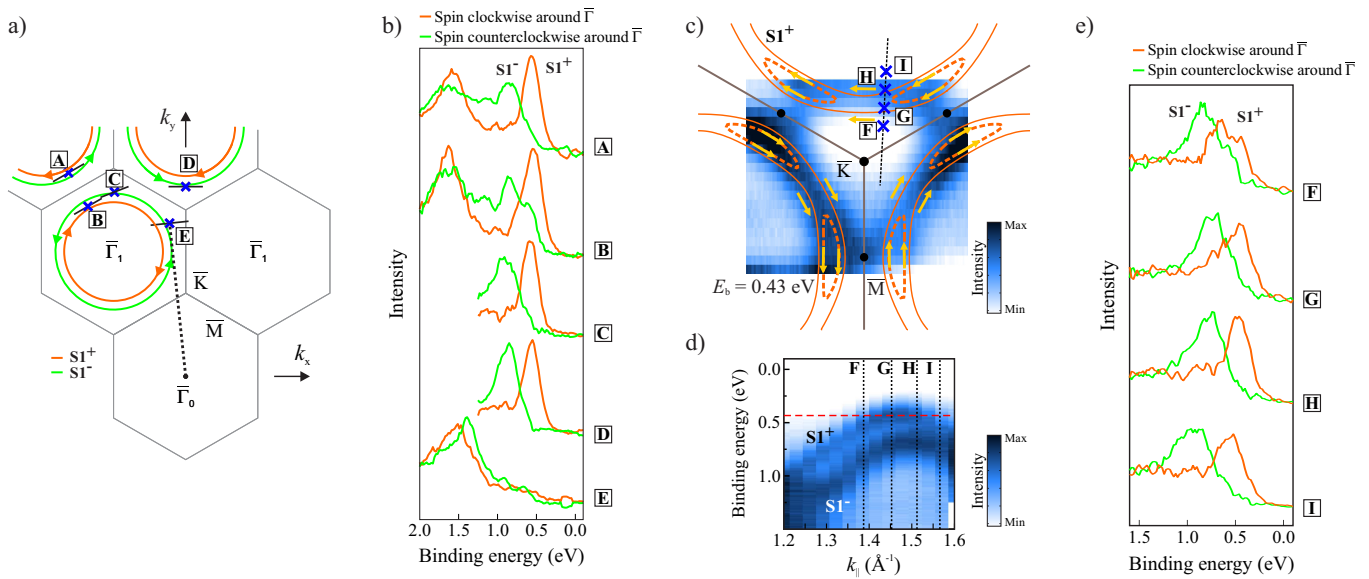}
\caption{a) Overview of points in $k$-space where spin-resolved PES measurements have been carried out. The orange and green lines show the overall vortical, Rashba-like, spin structure around \G of the \SP and \SN branches, respectively. b) Spin resolved EDCs from the points A--E in a). c) Experimental constant energy surface at $E_{\text{b}} = 0.43$~eV overlaid by calculated constant energy contours of the \SP branch from the TB model. Arrows show the calculated in-plane spin direction. Additional spin measurements have been performed in the points marked with blue crosses and labels F--I. d) Dispersion of the \SP and \SN branches along the dashed line in c). Horizontal dashed line is at $E_{\text{b}} = 0.43$~eV. e) Spin resolved EDCs at the positions F--I marked with vertical dashed lines in d) and blue crosses in c).}
\label{spinfig}
\end{figure*}
 
The aforementioned saddle points in the \SP and \SN branches have interesting consequences on the topology of the CE contours that develop when moving from the maximum energy of \SP and towards lower energies (direction of increasing binding energy in the experimental data). Initially, starting at the maximum of \SP, as we have already seen, CE contours in the shape of disconnected pockets develop outside \M --- indicated by the dashed contours in Fig.~\ref{spinfig}c). When moving further down in energy, these pockets grow larger until reaching the energy of the first saddle point of the \SP branch (SP$_1$) --- located outside \K. Exactly at this energy, the pockets have become so large that their endpoints touch. Going further down in energy, the pockets merge and two concentric contours appear, now enclosing \G, c.f. the solid contours in Fig.~\ref{spinfig}c). Since the disconnected pockets originate from the same branch, namely \SP, the spin does not circulate the contours, which explains the spin direction drawn as arrows in Fig.~\ref{spinfig}c). For the concentric contours that develop below the first saddle point, the in-plane spin direction should be the same for both the inner and the outer contour and circulate in the same direction as for the pockets since we are still only intersecting the \SP branch. Our calculations therefore confirm the predictions from ref.~\onlinecite{Frantzeskakis2010} of a parallel orientation of the spin of the inner and outer contours.

To experimentally verify that the in-plane spin component does not change direction between the inner and the outer solid contours drawn in Fig.~\ref{spinfig}c) we have performed spin-resolved ARPES measurements covering the maximum of the \SP branch along the \GK direction. The positions where the spin measurements have been performed are given by the crosses (with labels F--I) drawn in Fig.~\ref{spinfig}c) and the vertical dashed lines in Fig.~\ref{spinfig}d). The resulting spin-resolved energy distribution curves (EDCs) are presented in Fig.~\ref{spinfig}e). 

From the spin data, we immediately see that the spin direction on each side of the \SP maximum is the same and that it follows a clockwise circulation with respect to \G. This confirms the calculated spin structure discussed above. Additionally, we observe that the spin direction across the maximum of the \SN branch is unchanged and that it points in the opposite direction compared to that of \SP. That the spins of the \SP and \SN branches are opposite, and overall circulating around \G, we also see from the spin resolved EDCs shown in Fig.~\ref{spinfig}b). These EDCs are measured at the positions in $k$-space marked with letters A--E in Fig.~\ref{spinfig}a) and confirm that the spin of the \SP branch circulates clockwise around \G whereas for the \SN branch the rotation is counterclockwise. In Figs.~\ref{spinfig}a) and \ref{spinfig}b), the spin measurements are sensitive to the component of the spin in a direction perpendicular to the dashed line going from $\Gamma_0$ to the individual $k$-points where the measurements were performed. Consequently, in point E the spin is expected to have only a small component along this direction, something that is consistent with the spin resolved EDC for this point displayed in Fig.~\ref{spinfig}b). 

Our calculations also confirm that the \SP and \SN branches display a normal type of Rashba spin split, when approaching \G. Such a split of the S1 band is discernable in our data in Fig.~\ref{datafig}a) and \ref{datafig}b), although the data do not cover the close vicinity of \G. Thus, it appears that upon approaching \G, the S1 band behaves as expected for a Rashba system, in the same way as pointed out in ref.~\onlinecite{Sakamoto2009} for the S3 band. The split in the S3 band is also visible in our data, c.f. Fig.~\ref{datafig}a).
\begin{figure*}
\includegraphics[width=12 cm]{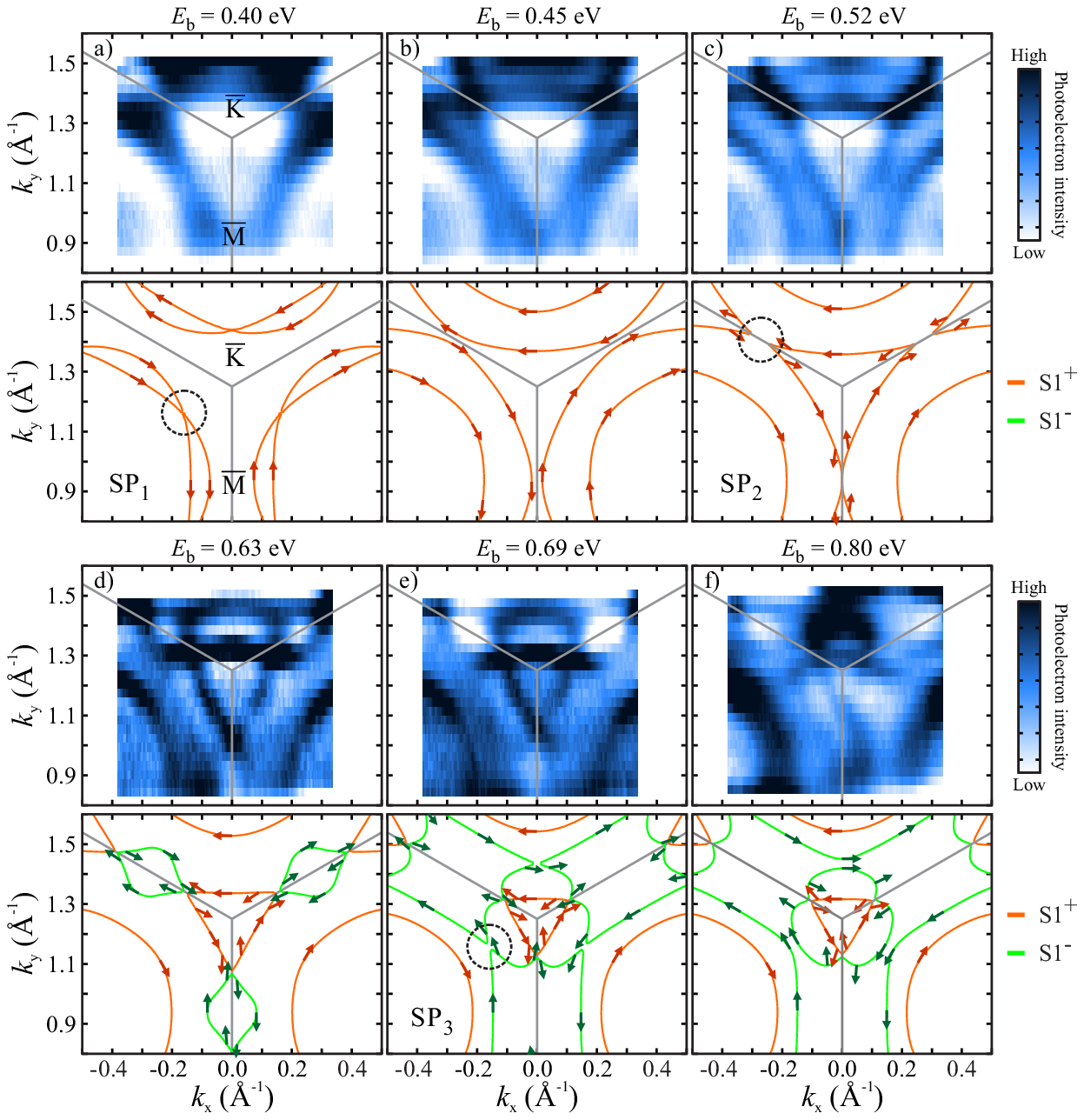}
\caption{Comparison of experimental constant energy surfaces around the \K-point with calculated contours. The given binding energies refer to the experimental data. The calculated energy scale have been adjusted to qualitatively match the experimental data. Orange lines represent contours for the \SP branch and green lines for the \SN branch. Arrows indicate the in-plane spin direction for the two branches, respectively. The panels labeled SP$_1$, SP$_2$ and SP$_3$ represent constant energy surfaces at, or very close to, the three saddle points discussed in the main text. Dashed circles indicate the location of the saddle points.}
\label{contours}
\end{figure*}

So far, we have only looked at what happens to the CE contours when crossing the energy of the uppermost (at lowest binding energy) saddle point of the \SP branch. As discussed earlier, two other saddle points exist in the S1 branches. In general, when traversing a saddle point along the energy direction, the topology of the CE contours will change. This is true also in our case, as shown by the different calculated CE contours in Fig.~\ref{contours}. The contours with a label SP$_x$, where $x=1,2,3$, indicates that the contours are taken at, or close to, the energy of the corresponding saddle point. In addition to the calculated contours, we show the corresponding experimental contours. The given binding energies refer to the experimental data. Arrows represent the calculated in-plane spin for points along the different contours.

From panels \ref{contours}a) and \ref{contours}b), we see what happens when passing through the uppermost saddle point (SP$_1$) of \SP. The disconnected pockets seen outside \M connect and concentric contours appear. When moving further down in energy (towards higher $E_{\text{b}}$) the concentric contours of \SP grow until the outer contour touches the BZ boundary at \M. This signals that the second saddle point (SP$_2$) of \SP has been reached. At this energy, \SP and \SN are degenerate at the \M-point. Decreasing the energy even further, panel \ref{contours}d), the topology of the contours belonging to the \SP branch change once more, now forming disconnected contours where one encloses \G and the others enclosing the \K points. Additionally, contours from the \SN branch start to develop around the \M-point. The latter continue to expand when moving down in energy until reaching the saddle point SP$_3$, which belongs to the \SN branch, see panel \ref{contours}e). Beyond this, the topology of the contours from \SN changes from separated contours around \M to one contour enclosing \G and others enclosing \K, consequently showing a similar topology as the \SP branch. 

Comparison between data and calculations in Fig.~\ref{contours} reveals that the modeled CE contours qualitatively agrees with the experimental observations. In particular, the behavior of \SP through the saddle points SP$_1$ and SP$_2$ is well captured by the model, as are the \SN pockets developing around \M and the flower-like structure of \SP and \SN around \K. 

Paying closer attention to the direction of the calculated in-plane spin, we can make two observations. Firstly, we see that the polarization of \SP and \SN are opposite, as confirmed by our spin measurements presented in Fig.~\ref{spinfig}b) and \ref{spinfig}e). As long as the constant energy contour one is looking at only cuts through the \SP branch, the spin direction, indicated by the arrows in Fig.~\ref{contours}, circulates in the same direction around \G for all contours, regardless of what the actual contours look like. In particular for the experimental contours at $E_{\text{b}} = 0.45$~eV, see Fig.~\ref{contours}b), the corresponding calculated contours give the same spin direction for the inner and outer lines. As we saw from Figs.~\ref{spinfig}c)--d), this behavior is confirmed by our spin-resolved measurements. This is in contrast to what is suggested by ref.~\onlinecite{Sakamoto2009}, where the two contours at the same binding energy are labeled with opposite spins. 

Secondly, and perhaps more interestingly, we observe that although both \SP and \SN form closed contours around \M and/or \K, the spin does not circulate these points in a continuous way as it does around \G. Looking, e.g., at panel d) in Fig.~\ref{contours}, the spin direction of both \SP and \SN contours make discrete flips when crossing the BZ boarder. In fact, it actually looks as if the \SN contour in one BZ is the natural continuation of the \SP contour in the neighboring zone. Connecting these contours at the BZ boundary leads to a smooth transition of the spin when crossing the boundary. However, the calculations yield an energy gap between \SP and \SN everywhere along the BZ boarder, i.e. along \KMK, except in the \M-point. As we saw earlier in Fig.~\ref{datafig}b), from our data, we cannot establish the existence of such an energy gap. Assuming that the gap yielded by the calculations is present, and the gap is sufficiently small, the spin in one branch would still be able to reverse its direction upon crossing the zone boundary if a non-zero Berry curvature exists, thus giving the appearance that the spin behavior across the zone boundary is continuous~\cite{Gradhand2012}. 

Regardless of the existence of this energy gap, we can make an important observation concerning the spin structure of the S1 band. The \SP and \SN branches have a more complicated dispersion compared to a simple Rashba model describing a free-electron like parabolic state. In spite of this, if we only consider the direction of the spin, we see that the deviations from what we expect for a parabolic state are small. Only close to the BZ boundary, the spin direction seems to deviate from a direction perpendicular to the $k$-vector. Along the BZ boundary \KMK the spins tend to align parallel to the zone boarder, and when approaching \K the spins seem to point slightly towards \K. The direction of the in-plane spin appears to be independent of what the actual dispersion of the band, and thus what the CE contours look like, and instead follows an overall circulating structure around \G as predicted by the standard Rashba model --- deviating only close to the zone boundary. Such a deviation suggest the presence of a non-zero Berry curvature, which can be interpreted as a local magnetic field in $k$-space that causes a rotation in the spins.

\begin{figure}%
\includegraphics[width=8.6 cm]{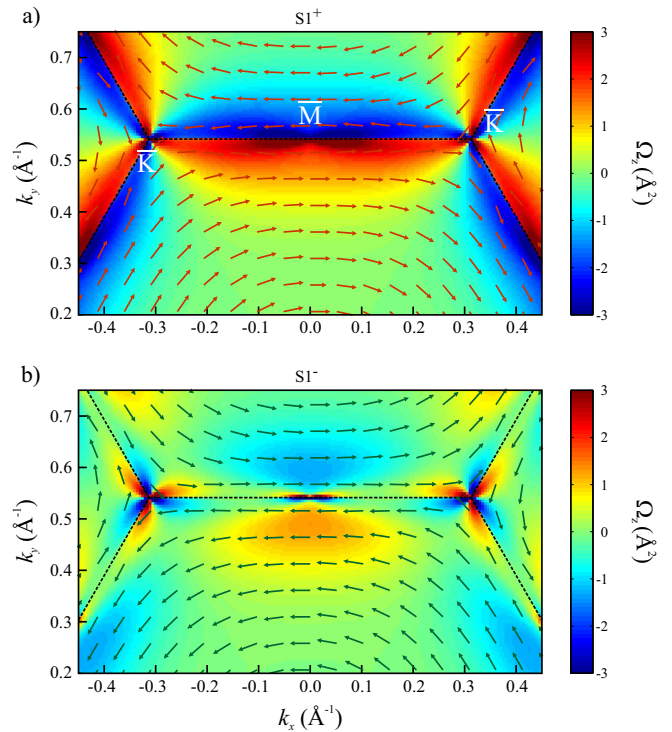}%
\caption{The $z$-component of the Berry curvature calculated using Eq.~(\ref{BCmain}) for the highest lying spin branches \SP/\SN (band at lowest binding energy). The calculated in-plane spin components are show as arrows. The overall spin behavior follows a vortical structure around \G which becomes hexagonally deformed for $k$-values approaching the BZ boundary. The local spin behavior around \K and \M follows naturally from the intersection of spin vortices centered around the \G-points in neighboring Brillouin zones.}%
\label{BerryCurvature}%
\end{figure}

In Fig.~\ref{BerryCurvature}, the Berry curvature for the \SP and \SN bands around \M and \K is shown together with the calculated in-plane spin component of the respective spin branches. One notes that the curvature has a rather complex structure around the high symmetry points. As seen from Eqs.~(\ref{BC1})-(\ref{BC2}), this is a result of the near degeneracies of the S1 band that occur close to these points. We see that in the regions with a non-zero Berry curvature, which occurs away from the BZ center and mostly when approaching the BZ boundary, the in-plane spins experience a stronger rotation away from a direction perpendicular to the $\bm{k}$-vector. It is also clear that the curvature has a three-fold rotational symmetry around the \G-point as expected from the atomic trimer surface structure. As pointed out, the overall spin behavior can be rather well understood as the result of a normal Rashba interaction. That this is so, is further confirmed by the fact that the Berry curvature oscillates in sign when going around all high symmetry points thus resulting in no net Berry phase --- consistent with what is expected for a topologically trivial system.  

\section{Summary and Conclusions}
In summary, we have used spin- and angle resolved PES in combination with TB band calculations to investigate the dispersion and the spin structure of the electronic surface states of the Bi/Si(111)-($\sqrt{3}\times\sqrt{3}$)R$30^{\circ}$ surface. Strictly speaking, the spin does not experience a vortex-like structure around \K. The band structure seems to suggest so, and when looking at the spin direction along \GK one could also believe this since the spin direction is tangential to the CE contours. However, when approaching the BZ boundary, the spins are significantly rotated away from a direction tangential to the constant energy contours enclosing \K and align semi-parallel to the BZ boundary in accordance with a vortical rotation around \G. 

The hole-like pocket developing from the \SN branch around \M also gives the impression that one has a spin vortex around this point. Again, by looking at the directions of the calculated spins, the spin remains largely parallel to the BZ boundary and hence shows that neither this point is the source of a spin vortex. In fact, the spin structure at \M is a natural result of the intersection of two spin vortices centered around \G points in neighboring Brillouin zones, as seen from Fig.~\ref{BerryCurvature}. In a similar way, the spin structure at \K arises naturally from the fact that the \K-point is the intersection of three BZs and thus three spin vortices from adjacent BZs, see Fig.~\ref{BerryCurvature}. Future detailed spin resolved measurements around the \K-point can be used to confirm that the spin is not tangential to the constant energy contours and thus does not circulate the \K-point.

The model we use for calculating the band structure and the in-plane spin components does not need any peculiar effect to explain what we observe in the experiment. The threefold symmetry of the calculated Berry curvature tells us that there is no net curvature when circulating the high symmetry points of the BZ, thus confirming the topologically trivial nature of this system. 

Although the surface-state band structure of the specific system studied here is more complex in comparison with the free-electron like states observed in the model systems used to demonstrate the Rashba effect, e.g. surfaces of noble metals, we believe that the spin behavior in our system is inherently determined by the standard Rashba effect. The reason for the seemingly complex spin structure around the \K points and the hexagonal deformation from a pure vortex is a result of having an in-plane spin vortex confined to a hexagonal Brillouin zone. We speculate that this is a general property of topologically trivial Rashba systems hosting spin-split states for which the atomic structure results in a hexagonal BZ. However, to experimentally access the spin structure in the vicinity of the symmetry points on the BZ boundary, one requires systems where the Fermi level is placed such that the spin branches are populated all the way to the edge of the BZ. 

We believe that our work will pave the way for future studies of the universal spin behavior of Rashba systems with a hexagonal symmetry. 

\section{Acknowledgements}
This work was made possible through support from the Knut and Alice Wallenberg Foundation and the Swedish Research Council.

%------------------------------------ REFRERENCES
%\bibliography{references}
%merlin.mbs apsrev4-1.bst 2010-07-25 4.21a (PWD, AO, DPC) hacked
%Control: key (0)
%Control: author (8) initials jnrlst
%Control: editor formatted (1) identically to author
%Control: production of article title (-1) disabled
%Control: page (0) single
%Control: year (1) truncated
%Control: production of eprint (0) enabled
%

\end{document}